\documentclass[%
reprint,
 amsmath,amssymb,
 aps,
]{revtex4-1}

\usepackage[dvipdfmx]{graphicx}
\usepackage{bm}

\begin{document}

\preprint{APS/123-QED}

\title{Quantum interferometric generation of polarization entangled photons}

\author{Haruka Terashima}
\author{Satoshi Kobayashi}
\author{Takaho Tsubakiyama}%
\author{Kaoru Sanaka}%
\affiliation{Department of Physics, Tokyo University of Science, Shinjuku-ku, Tokyo 162-8601, Japan\\}%

\begin{abstract}
Quantum interference, like Hong-Ou-Mandel interference, has played an important role to test fundamental concepts in 
quantum physics. We experimentally show that the multiple quantum interference effects enable the generation of high-performance 
polarization entangled photons. These photons have a high-emission rate, are degenerate, have a broadband distribution, and are postselection free. A quantum interferometric scheme, based on a round-trip configuration of a double-pass polarization Sagnac interferometer, makes it possible to use the large generation efficiency of polarization entangled photons in the process of parametric down-conversion and to separate degenerate photon pairs into different optical modes with no requirement of postselection. We demonstrate experimentally that multiple quantum interference is not only an interesting fundamental quantum optical phenomenon but can be used for novel photonic quantum information technologies. 
\end{abstract}

\pacs{Valid PACS appear here}
\maketitle


\section*{Introduction}

Hong-Ou-Mandel (HOM) interference is typical quantum interference which was originally used to observe the bosonic nature of photons \cite{hong87}. HOM and more generalized quantum interference still play an important role in fundamental quantum optics\cite{kim15}. HOM interference is also useful for verifying the indistinguishability of independently generated single photons emitted from remote atomic systems\cite{beugnon06,sanaka09,flagg10,lettow10,patel10,sipahigil12}. Furthermore, the HOM effect has potential applications in photonic quantum information technologies particularly as a method of creating entangled photons for quantum cryptography\cite{ekert91}, quantum dense coding\cite{mattle96}, quantum repeater\cite{pan98}, and quantum teleportation\cite{bouwmeester97}. Theoretically, it is possible to generate unconditional entangled photons by using HOM interference effects and huge nonlinear interactions\cite{chuang95} or  effective nonlinear interactions using linear optical elements\cite{knill01}. However it has been very difficult to experimentally achieve the necessary nonlinear interaction at the single-photon level\cite{turchette95,sanaka04}. Instead, spontaneous parametric down-conversion  (SPDC) in a second-order nonlinear process has been used to directly produce polarization entangled photon pairs. A general approach uses a thin $\beta$-barium borate (BBO) bulk crystal with type-II phase matching \cite{kwiat95} or sandwiched BBO crystals with type-I phase matching to generate noncollinearly propagating polarization entangled photon pairs \cite{kwiat99}. Owing to the development of quasi-phase matching techniques, the periodically poled KTiOPO$_{4}$ (ppKTP) or LiNbO$_{3}$(ppLN) also has become a standard technique\cite{armstrong96}. Several types of entanglement sources have been developed by combining such nonlinear crystals and interferometric configurations such as Mach-Zehnder\cite{fiorentino04}, Franson\cite{sanaka02}, and Sagnac interferometers\cite{shi04,kim06,steinlechner14}.

Among the different interferometric schemes, the Sagnac interferometer has major advantages because the symmetric geometry of the setup allows a very phase-stable condition resulting in the generation of high-quality polarization entangled photons. In particular, the scheme with orthogonally polarized photon pairs obtained by type-II SPDC makes it possible to separate degenerate polarization entangled photon pairs into different optical modes with no postselective detection\cite{kim06}. The scheme with ordinary type-0 or type-I SPDC requires a nonpolarizing beam splitter to separate degenerate photon pairs with a 50$\%$ probability of success\cite{shi04}, wavelength filtering to separate nondegenerate pairs of photons\cite{steinlechner14}, or spatial mode filtering to separate non-collinear down-converted photons\cite{jabir}. In contrast, a double-pass configuration with type-0 or type-I SPDC has the major advantages of a simple setup and a high emission ratio of photon pairs\cite{steinlechner13}. Bi-directional pumping to a single ppKTP crystal generates polarization-entangled photons from two sets of parallel polarized photon pairs on the collinear optical mode. A non-polarizing beam splitter or color filters are necessary to separate polarization-entangled photon pairs conditionally. 

In this paper, we present a quantum interferometric scheme to generate polarization entangled photons that satisfies the properties of both type-0 and type-II SPDC simultaneously by the integration of interferometric and double-pass configurations.  Our approach uses a multiple reverse process of HOM interference. The multiple quantum interference effect makes it possible to use the largest second-order nonlinear coefficient of a nonlinear crystal to generate polarization entangled photons and also to separate degenerate photon pairs into different optical modes with a 100$\%$ probability of success in principle. 


\section*{Theory}

The principle of operation of our scheme is shown in Fig.1. The polarization of the pump beam is represented by the amplitudes of its horizontal (H) and vertical (V) components and the relative phase correlation. The laser is injected into a polarization Sagnac interferometer consisting of a polarizing beam splitter (PBS), two half-wave plates set at 45$^{\rm o}$ (HWP1) and 22.5$^{\rm o}$(HWP2) ,  a ppKTP crystal, and mirrors. The polarization optics with the setup operate for both wavelengths of the pump laser field and the down-converted photons. 

\begin{figure}
 \centering
 \includegraphics[keepaspectratio, scale=0.7]
      {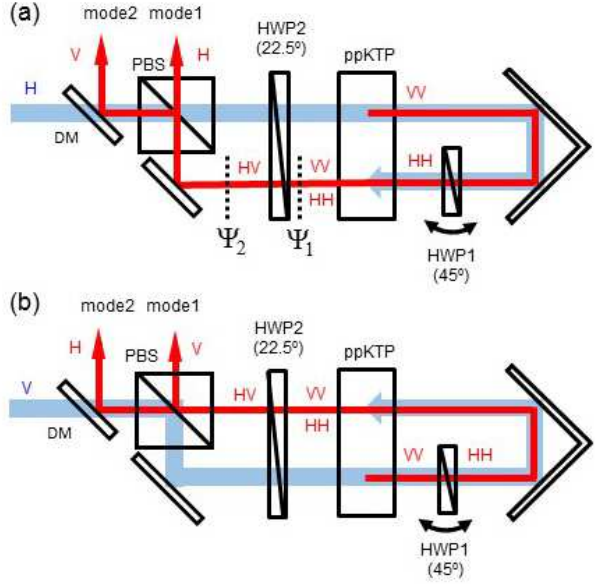}
 \caption{ (Color online) Schematic of integrated double-pass polarization Sagnac interferometer for (a) clockwise and (b) counterclockwise directions. The polarizing beam splitter (PBS) and half-wave plates (HWP1, HWP2) operate for both wavelengths of the laser beam and the down-converted photons. A dichroic mirror (DM) separates the pump laser and down-converted photons. Blue and red characters show the horizontal (H) or vertical (V) polarization state of the pump laser and down-converted photons, respectively. }
 \label{fig1}
\end{figure}

The H-component of the pump laser passes through the PBS as shown in Fig.1(a) and round trips the setup in the clockwise direction. The polarization of the pump laser is inverted into 45$^{\rm o}$ i.e. the superposition state of the H- and V-polarization states through HWP2.  Here we suppose that the V-component of the pump laser generates the V-polarized photon pairs under type-0 SPDC. These down-converted photon pairs are inverted to become H-polarized by HWP1 set at 45$^{\rm o}$, and injected into the ppKTP crystal again with the pump laser beam. The photon pairs generated by the second SPDC are V-polarized and superposed with the H-polarized photon pairs generated by the first SPDC in a collinear optical mode. The polarization state of the photon pairs after the second SPDC can be expressed as:    
\begin{equation}
 | \Psi_{\rm 1} \rangle = \frac{1}{2}  [ \left( \hat a_H^\dag \right)^2 + \mathrm{e}^{i\phi} \left( \hat a_V^\dag \right)^2 ] | 0 \rangle,
\label{psi_h1}
\end{equation}
where $\hat a_H^\dag$ and $\hat a_V^\dag$ are the creation operators for H- and V-polarized down-converted photons respectively.
The angle of HWP1 setting at 45$^{\rm o}$ results in the generation of H- and V-polarized photon pairs with same amplitude as shown by Eq. (\ref{psi_h1}). Here $\phi$ is the relative phase between the photon pairs from the first and second SPDCs, and $ | 0 \rangle$ is the vacuum state. The phase does not vary in time because it is caused by HWP1's material dispersion between the pump laser and the down-converted photons, and is fully adjustable by tilting HWP1. These operators are unitarily transformation by HWP2 (set at 22.5$^{\rm o}$) as $ \hat a_H^\dag \rightarrow (  \hat a_H^\dag + \hat a_V^\dag )/\sqrt{2}$ and $ \hat a_V^\dag \rightarrow ( -\hat a_H^\dag + \hat a_V^\dag )/\sqrt{2}$. The polarization states of the photon pairs output from HWP2 can be expressed as 
\begin{equation}
  | \Psi_{\rm 2} \rangle  = \frac{e^{i \frac{\phi}{2}}}{2}  [ \cos \frac{\phi}{2} \cdot \{  \left( \hat a_H^\dag \right)^2 + \left( \hat a_V^\dag \right)^2 \}  -2i \sin \frac{\phi}{2}  \hat a_H^\dag \hat a_V^\dag ] | 0 \rangle.
\label{psi_h2}
\end{equation}
When the phase is set at $\phi = \pi$ by tilting HWP1, the state $| \Psi_{\rm 2}  \rangle $ becomes $ \hat a_H^\dag \hat a_V^\dag | 0 \rangle \equiv |1_H,  1_V \rangle $. Under the phase condition, the whole unitary transformation can be represented as
\begin{equation}
\frac{1}{\sqrt{2}} ( | 2_H \rangle - | 2_V \rangle) \rightarrow |1_H,  1_V \rangle. 
\label{HOM}
\end{equation}
This process is the quantum interference corresponding to the reverse process of HOM interference on polarization basis \cite{hong87}. Since H-photons pass through the PBS and V-photons are reflected by the PBS, the polarization state of the photon pairs output from the PBS is represented as $ | 1_H \rangle_1| 1_V \rangle_2 $  for optical modes 1 and 2 in Fig.1. 

On the other hand, the V-component of the pump laser is reflected by the PBS, as shown in Fig.1(b), and round trip in the counterclockwise direction. Through a similar process to multiple type-0 SPDCs and unitary transformations, the polarization state of the output from the PBS becomes $e^{i \theta} | 1_V \rangle_1| 1_H \rangle_2 $, where $\theta$ is the relative phase between  photon pairs generated from the clockwise and counterclockwise directions originally determined by the relative phase between the H- and V-components of the pump laser. 

When the H- and V-components of the pump laser are balanced, the outputs of photons generated from the clockwise and counterclockwise directions are also balanced. The photon pairs generated from the clockwise and counterclockwise directions are superposed on  output modes 1 and 2 with the same amplitudes. Upon normalization, the output state can be represented as
\begin{equation}
  | \Psi_{\rm OUT} \rangle =\frac{1}{\sqrt{2}} ( | 1_H \rangle_1 | 1_V \rangle_2 + \mathrm{e}^{i\theta} | 1_V \rangle_1 | 1_H \rangle_2).
\label{psi_out}
\end{equation}
Since $\theta$ is determined by the relative phase between the H- and V-components of the pump laser,  $\theta$ can be set to an arbitrary value by simply preparing appropriate polarization optics for the laser. When the pump laser is prepared with a 45$^{\rm o}$ linear polarization state,  the H- and V-components of the pump laser are balanced and the relative phase becomes $\theta=0$.  The output state becomes one of the Bell states $ \Psi^{+} $ through multiple two-photon quantum interference. Our scheme makes it possible to use the largest second-order nonlinear coefficient of a nonlinear crystal to generate photon pairs with no need for postselection.

\section*{Experiment}

Fig.2 shows the overall experimental setup. A 405 nm grating-stabilized single-frequency laser diode (LD) is coupled to a singlemode fiber. The output laser from the fiber with a power of about 540 $\mu$W passes through an isolator, a half-wave plate (HWP), a quarter-wave plate (QWP), a 300 mm focus lens, and a short-pass dichroic mirror (DM), and enters the Sagnac interferometer shown in Fig.1. These polarization optics in the interferometer operate for the wavelengths of laser (405 nm) and that of the down-converted photons (810nm). The crystal is mounted on a temperature controller and stabilized to a temperature of $27.7 \pm 0.1 \ ^{\rm o}$C. Owing to the setup and the self compensation property of the Sagnac interferometer  for path-length drifting, the setup retains almost the same phase-stable condition and the same interference visibility over several days of measurements. 
\begin{figure}
 \centering
 \includegraphics[keepaspectratio, scale=0.5]
      {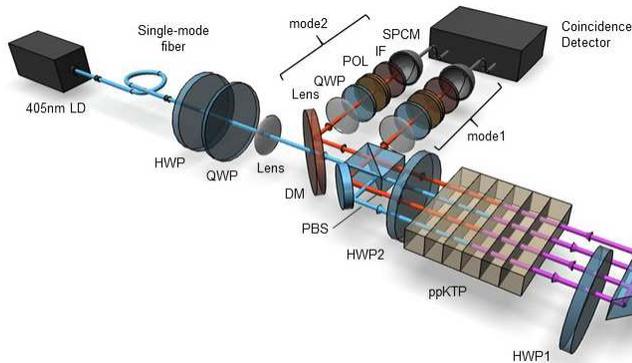}
 \caption{  (Color online) Overall experimental setup for generating polarization entangled photons through multiple quantum interference of type-0 parametric down-converted photons. The polarization state of the pump beam from the laser diode (LD) is set by the first half-wave plate (HWP) and quarter-wave plate (QWP) to adjust the relative amplitude and relative phase of the clockwise and counterclockwise outputs. The output photons are pass through lenses, QWPs, polarizers (POLs), and interference filters (IFs) in modes 1 and 2, and are detected by single-photon counting modules (SPCM). }
 \label{fig2}
\end{figure}

The poling period of the ppKTP crystal is 3.425 $\mu$m which is designed for collinear type-0 SPDC with 405 nm laser pumping and the crystal has an antireflection coating for both wavelengths. The ppKTP crystal in the interferometer is 10 mm long (crystallographic  $x$ axis), 10 mm wide ($y$ axis), and 1 mm thick ($z$ axis).  

The output photons for the two directions are collected with 300 mm focus lenses, then passed through QWPs, polarizers (POLs), interference filters (IFs) with a 810 nm center and 3 nm band width, and coupled to multimode fibers for detection. After detecting the photons with single-photon counting modules (SPCMs) constructed from Si avalanche photodiodes, the polarization states of the photon pairs  are analyzed with a coincidence detector consisting of a multichannel analyzer and a time-to-amplitude converter with a coincidence window about 2.3 ns. We remove accidental coincidence counts from the raw data and show pure coincidences in the following data.

\section*{Results}

Figs.3(a) and (b) show the coincidence counts for generated photon pairs set in the $\Psi^{+} $ state under a linear polarization basis  as a function of the mode 1 POL angle when the mode 2 POL is fixed at 0$^{\rm o}$ and 45$^{\rm o}$ respectively. We remove the QWPs in modes 1 and 2 for the measurement. The solid curves are best sinusoidal fits to the corresponding data. The visibility of the fringe is defined by $(C_{\rm max}-C_{\rm min})/(C_{\rm max}+C_{\rm min}) $, where $C_{\rm max}$ is the maximum coincidence count and $C_{\rm min}$ is the minimum coincidence count. The visibilities estimated from the sinusoidal fits in Figs.3(a) and (b) are $0. 98\pm0.02$ and $0. 83\pm0.02$ respectively. Fig.3(c) shows the coincidence count for the same state under a circular polarization basis as a function of the mode 1 POL angle when the mode 2 POL angle is fixed at 45$^{\rm o}$ and the angles of both QWPs are set at 0$^{\rm o}$. The estimated visibility is $0. 80\pm0.02$.  These reasonable visibilities are clear evidence of nonclassical quantum interference given by the polarization entangled state. These visibilities are currently limited, mainly because we use multimode fibers to collect the generated entangled photons since the tilting of HWP1 affects for the overlapping of spatial modes.  The birefringence effect of ppKTP crystal also affects the mode mismatch between the photons from first and second SPDC. It will be possible to improve the visibilities by using single spatial mode filters and additional compensation crystals in the future. 
\begin{figure}
 \centering
 \includegraphics[keepaspectratio, scale=0.6]
      {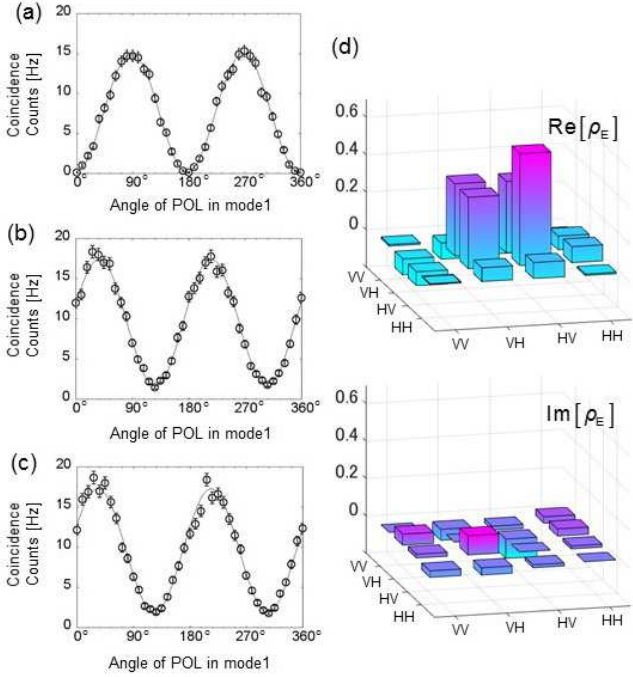}
 \caption{(Color online) Plot of coincidence counts under linear polarization basis when mode 2 POL is fixed at  (a) 0$^{\rm o}$ and  (b) 45$^{\rm o}$ and (c) circular polarization basis. Solid lines are best sinusoidal fits to data. (d) Reconstructed two-photon density matrix with real and imaginary parts obtained by quantum state tomography.}
 \label{fig3}
\end{figure}

As a nonlocality test of the entanglement, we measure the $S$ parameter in the Clauser-Horne-Shimony-Holt form of Bell's inequality \cite{clauser69}. Following the method described in \cite{kwiat95,kim06}, we obtain $S=2.59\pm0.05$, which corresponds to the violation of the classical limit of 2 by more than 17 standard deviations. 

We also measure the photon pairs by reconstruction of the complete density matrix by means of quantum state tomography, which requires coincidence measurements for 16 combinations of polarization bases\cite{altepeter05}. The experimentally reconstructed real and imaginary parts of the two-photon polarization density matrix $\rho_\mathrm{E}$ are shown in Fig.3(d). The fidelity of the experimentally reconstructed density matrix to an ideal entangled state $\Psi^{+} $ is given by $F=\langle \Psi^{+} | \rho_\mathrm{E}| \Psi^{+}  \rangle$. We estimate the fidelity as $0.873 \pm 0.003$ which is sufficiently high to show that the generated photon pairs are in the $\Psi^{+} $ state. 

Fig.4(a) shows the single counts (open circles) and coincidence counts (solid circles) of type-0 down-converted photons directly obtained from the same ppKTP crystal. From the data, the photon-pair production rate directly from the crystal is $N_{\rm pair} =5.86\pm0.05\times10^7$Hz/mW, which is reasonable compared with the rate under the actual round-trip configuration shown in Fig.2 of 540 $\mu$W pump power with $N_{\rm pair} \sim1.3\times10^7$ Hz/mW.  The actual photon-pair output rate estimated from the photon coupling efficiencies is $N_{\rm out}\sim 3.6 \times 10^7$ Hz/mW. These numbers are also comparable to the photon-pair production rate for a similar high-brightness degenerate entangled source of $N_{\rm pair} \sim3.9\times10^7 $ Hz/mW \cite{jabir}. We expect that it will be possible to improve these rates and efficiencies using waveguide structures and single-mode fiber settings in future experiments. 
\begin{figure}
 \centering
 \includegraphics[keepaspectratio, scale=0.8]
      {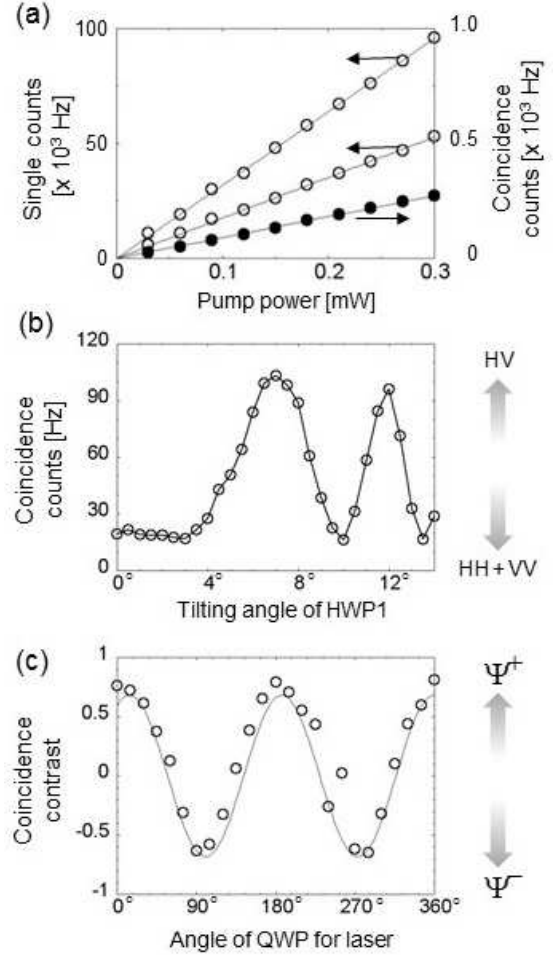}
 \caption{ (a) Plot of single and coincidence counts of type-0 down-converted photons directly obtained from the ppKTP crystal as a function of input pump power with filters of 3 nm band width. (b) Plot of coincidence counts corresponding to measured photon pairs produced by the clockwise process for different dispersion effects depending on the angle of the HWP1. (c) Plot of polarization correlation to show the switching of Bell states between $\Psi^{+}$ and $\Psi^{-}$ depending on the angle of the QWP for the laser.}
 \label{fig4}
\end{figure}

Fig.4(b) shows the coincidence counts of the measured photon pairs generated through an optical setup that corresponds to the clockwise process as a function of the tilting angle of HWP1 shown in Fig.1(a). By tilting the angle of HWP1, the dispersion effect between the pump laser and the down-conversion changes $\phi$ in Eq. (\ref{psi_h2}). For the measurement of coincidences given by photon pairs with the state $ | 1_H \rangle_1| 1_V \rangle_2 $, the polarization state of the pump laser is set to horizontal. We remove the QWPs and POLs in modes 1 and 2 for the measurement. The coincidence clearly depends on the tilting angle and verifies the operation in the clockwise direction described by Eq. (\ref{psi_h2}). 

We also carry out a measurement to show the flexibility of our scheme in terms of its ability to prepare all four Bell states. Fig.4(c) shows the measured polarization correlation as a function of the QWP angle for each laser polarization setting. The polarization correlation is defined by $(C_{DD}-C_{DA})/(C_{DD}+C_{DA}) $, where $C_{DD}$ ($C_{DA}$) is the coincidence count when the POL in mode 1 is set at  45$^{\rm o}$, and the POL in mode 2 is set at 45$^{\rm o}$ (-45$^{\rm o}$). We remove the QWPs in modes 1 and 2 for the measurement. Owing to the phase correlation, the state $\Psi^{+} $ ($\Psi^{-}$) gives the maximum  $C_{DD}$ ($C_{DA}$) and minimum  $C_{DA}$ ($C_{DD}$ ). The solid curves are best sinusoidal fits to the corresponding data. Because the setting of the QWP for the laser adjusts the relative phase $\theta$ in Eq. (\ref{psi_out}), the contrast clearly depends on the angle. The results show that our setup has the ability to easily switch the Bell state between  $\Psi^{+} $ and $ \Psi^{-} $. An additional HWP in an output mode can be used to convert photon pairs from states $\Psi^{+} $ and $\Psi^{-} $ into the other two Bell states $\Phi^{+} $ and $\Phi^{-}$,  respectively \cite{kwiat95}.  \\

\section*{Discussion}

We compare the spectral characteristics and photon-pair generation efficiency between type-0 and type-II SPDC. For collinear propagation along the crystallographic $x$ axis, KTP has three nonzero second-order tensor coefficients $d_{ijk}$\cite{pack04}. The nonlinear coefficient of KTP for the type-II process is $d_{yyz}\sim3.9$ pm/V. The largest nonlinear coefficient of KTP of $d_{zzz}\sim18.5$ pm/V is associated with the type-0 process, which is about 4.7 times larger than that for the type-II process\cite{steinlechner14}. Since photon-pair generation efficiency is given by the square of  the second-order tensor coefficients, we expect that the factor for type-0 SPDC is  $(d_{zzz}/d_{yyz})^2 \sim 22.5$ times larger than that for type-II SPDC.  

Steinlechner $et\ al.$ reported that the count rate of polarization entangled photon pairs for nondegenerate type-0 SPDC is about 10 times higher than that for degenerate type-II SPDC per unit bandwidth and per unit pump power. The total photon-pair production rate per unit pump power is two orders of magnitude higher owing to the large bandwidth of type-0 SPDC\cite{steinlechner14}. It is reasonable to expect a similar or higher production rate when both type-0 SPDC and type-II SPDC are degenerate. Recently, a source of entangled photons with such a large bandwidth has attracted considerable attention for use in quantum optical coherence tomography\cite{okano05}. A large bandwidth of correlated photon pairs gives a very short coincidence timing owing to the relation used in the Fourier transformation. This property also has several potential applications such as achieving ultrashort temporal correlations through nonlinear interactions with the flux of entangled photons\cite{dayan05}, metrology methods using the very narrow dip in HOM interference\cite{nasr08}, quantum clock synchronization\cite{giovannetti01}, time-frequency entanglement measurement\cite{hofmann08}, and multimode frequency entanglement\cite{mikhailova08}.

Owing to the simplicity of our scheme, it is possible to further improve the efficiency of polarization entangled photons by pulse laser pumping and the fabrication of waveguide structures on nonlinear crystals\cite{fiorentino07,levine11}. Compensation crystals are essential because the walk off effects in the crystals when we use pulse laser pumping. It is also possible to generate photons in the telecom-band wavelength region by selecting the poling period of the ppKTP crystal\cite{jin14}. The method of using multiple quantum interference is useful not only for generating postselection-free polarization entangled photons but also for the further amplification of entangled photons through the stimulated emission of SPDC\cite{lamas01}. 

\

\section*{Summary}

We experimentally demonstrated the generation of high-performance polarization entangled photons through multiple reverse processes of HOM interference. The quantum interferometric scheme allows not only the use of the largest second-order nonlinear coefficient for the generation of polarization entangled photons in the process of parametric down-conversion, but also the separation of degenerate photon pairs into different optical modes with no requirement of postselective detection. Through experiments, we showed that the multiple quantum interference effect is interesting not only from the view point of fundamental quantum optics but also as a method for generating high-quality polarization entangled photons for novel quantum information technologies. \\ \\ \\ \\ \\ \\ \\

\section*{Methods}

To estimate the photon-pair production rate with our source, we directly measure the pump power dependence of the single and coincidence counts for type-0 down-converted photons using the same ppKTP crystal. The focus and collimation conditions are similar to those in the setup in Fig.2 using same lenses. The down-converted photons are split into two optical modes using a 50$\%$-50$\%$ nonpolarizing beam splitter (BS) and detected under the same DM and IF with 810 nm center and 3 nm band width. The photon-pair production rate is defined as
\begin{equation}
N_{\rm pair}=\frac{N_1\times N_2}{N_C},
\label{pppr}
\end{equation}
where  $N_1$ ($N_2$) is the single count rate per unit input power in mode 1 (2) and  $N_C$ is the coincidence count rate per unit input power \cite{tanzilli01}. The slopes of the plotted data in Fig.4(a) give $N_1=3.129\pm0.014\times10^5$ Hz/mW, $N_2=1.698\pm0.009\times10^5$ Hz/mW, and  $N_C=9.065\pm0.054\times10^2$ Hz/mW. Therefore the rate of direct photon-pair production from the crystal is $N_{\rm pair} =5.86\pm0.05\times10^7$ Hz/mW. The actual photon-pair output rate estimated from the photon coupling efficiencies is given by  
\begin{equation}
N_{\rm out}=\frac{N_C}{\eta_1\times \eta_2},
\label{nout}
\end{equation}
\clearpage
where  $\eta_1$ ($\eta_2$) is the total coupling efficiency determined by the photon transmission and detection efficiency. Since the FWHM bandwidth of type-0 SPDC under the degenerate condition is $\sim 30$ nm, the photon transmission efficiency through a 3nm bandwidth IF is typically $\eta_{\rm IF} \sim 0.1$. The transmission efficiency through a BS is typically $\eta_{\rm BS} \sim 0.5$. The coupling efficiency determined by unavoidable optical and detection losses is typically  $\eta_{\rm D} \sim 0.1$. Therefore the total coupling efficiency is  $\eta_{1,2} =\eta_{\rm IF} \times \eta_{\rm BS} \times \eta_{\rm D} \sim 0.005$. Therefore the actual photon-pair output rate estimated from the photon coupling efficiencies is $N_{\rm out}\sim 3.6 \times 10^7$Hz/mW.


\end{document}